\documentclass[twocolumn,amsmath,amssymb]{revtex4}

\usepackage{graphicx}
\usepackage{dcolumn}
\usepackage{bm}

\begin{document}

\title{Schwinger-boson calculation for a frustrated antiferromagnet}
\author{Michael J. Quist}
\email{mjq1@cornell.edu}
\affiliation{Department of Physics, Cornell University, Ithaca, New York 14850}

\date{\today}

\begin{abstract}
The application of the Schwinger-boson transformation to quantum Heisenberg
magnets is briefly reviewed, beginning with the derivation of a rotationally invariant
mean-field theory.  The inclusion of Gaussian fluctuations is discussed in some
detail for a general (non-Bravais) lattice, extending available results for simple
lattices.  Numerical results are presented for the ground-state energy of the
Shastry-Sutherland model, and possible future refinements of the method are
outlined.
\end{abstract}

\maketitle

\section{Introduction}

The Schwinger-boson method is a versatile means of studying quantum 
spin systems~\cite{arovas,aa,auerbach}.  It relies on the 
correspondence between spin and boson degrees of freedom
defined by the Schwinger-boson transformation.  The introduction
of fictitious boson degrees of freedom permits the derivation of
simple mean-field theories, with rotationally invariant order parameters
which cannot be defined in terms of the original spins.
Because the procedure explicitly preserves rotational invariance, 
spontaneous symmetry breaking need not be assumed, in contrast to 
standard spin-wave approaches.  The method is therefore capable of
describing collinear, spiral, and spin-liquid phases within a unified framework,
where broken-symmetry phases are represented by Bose-condensed
phases of Schwinger bosons~\cite{hirsch}.
At the mean-field level, early Schwinger-boson calculations produced 
quantitatively disappointing results.
However, small changes in the formalism greatly improve its
accuracy without unduly complicating its use~\cite{twist1}.
In this paper we briefly review the application of the Schwinger-boson
transformation to quantum Heisenberg magnets, beginning with the derivation of
a rotationally invariant mean-field theory.
We then show in some detail how Gaussian fluctuations around
the mean-field solution can be included for arbitrary (non-Bravais) 
lattices.  This generalizes the available results, which treat
simple lattices.  As an application, we examine the Shastry-Sutherland
model, which has been the subject of renewed interest due to its
experimental realization in $\rm{SrCu_{2}(BO_{3})_{2}}$~\cite{expt1}.

\section{Schwinger-boson method}

The Schwinger-boson transformation is an exact mapping from spin to boson
degrees of freedom, given by
$\hat{\bf S} \mapsto \frac{1}{2}
({\hat{a}}^{\dag} \, {\hat{b}}^{\dag})_{m}
{\bm\sigma}_{mn}
(\hat{a} \, \hat{b})_{n}$;
here $m,n=1,2$, and ${\bm\sigma}$ is the vector 
of Pauli matrices.  The original spin Hilbert space corresponds
not to the full boson Hilbert space but rather to the 
($2S+1$)-dimensional subspace in
which
$
{\hat{\delta n}} = 
{\hat{a}}^{\dag}{\hat{a}}+{\hat{b}}^{\dag}{\hat{b}}-2S = 
0
$.  Our goal is to evaluate the partition function,
$
{\cal Z}(\beta) = {\rm Tr}_{\rm phys}\,e^{-\beta\hat{\cal H}}
$, where
$
{\hat{\cal H}}(\{ \hat{\bf S}_{i} \}) \mapsto
{\hat{\cal H}}(\{ \hat{a}_{i}, \hat{b}_{i} \})
$
is an arbitrary spin Hamiltonian transformed to a boson Hamiltonian, and the trace is 
restricted to the physical subspace.  We can replace the restricted
trace with a trace over the full boson Hilbert space using an
appropriate projection operator for each spin:
\begin{equation}
\hat{\cal P}_{i} = \int_{-\pi}^{\pi} \frac{d\lambda_{i}}{2\pi}
\exp{\left(-i \lambda_{i} \hat{\delta n}_{i} \right)}.
\end{equation}
Then ${\cal Z}={\rm Tr}\,(\prod_{i}\hat{\cal P}_{i}) e^{-\beta \hat{\cal H}}$,
which can be rewritten as a functional integral in the usual way.  In 
particular, we obtain
\begin{equation}
\label{functionalZ1}
{\cal Z}(\beta) = 
\frac
{
\int{\cal D}\phi_{a}{\cal D}\phi_{b}{\cal D}\lambda
\,\,
\delta[\partial_{\tau}\lambda]
\exp\left(
-{\cal A}[\phi_{a},\,\phi_{b},\,\lambda]\right)
}
{
\int{\cal D}\lambda\,\,\delta[\partial_{\tau}\lambda]
},
\end{equation}
where the Euclidean action is
\begin{equation}
\label{action1}
{\cal A}
=
\int_{0}^{\beta} \!\!\! d\tau  
\left(
{\cal H}(\{\phi_{a i}, 
\phi_{b i}\}) + \sum_{i}i\lambda_{i}{\delta n}_{i}
+ \!\!\! \sum_{i;\,s=a,b} \!\!\! \phi_{s i}^{*}\partial_{\tau}\phi_{s i}
\right).
\end{equation}
The Lagrange multipliers $\{\lambda_{i}\}$ have been promoted to 
time-dependent variables; the inclusion of 
$
\delta[\partial_{\tau}\lambda]
\equiv
\prod_{i,\tau}\delta(\partial_{\tau}\lambda_{i})
$ in the functional measure
ensures that ${\cal Z}$ remains the same.

The action (\ref{action1}) is invariant under local gauge transformations 
parametrized by $\{\theta_{i}(\tau)\}$: $\lambda \rightarrow \lambda - 
\partial_{\tau}\theta, \, \phi \rightarrow e^{i\theta}\phi$.
This gauge symmetry is associated with the conservation of total boson 
number on each site, which holds simply because the magnitude
of each spin is constant.  The gauge is partially fixed in Eq.~(\ref{functionalZ1}) by
the condition that 
$\partial_{\tau}\lambda=0$; the remaining, unfixed, transformations are those
where $\theta$ is time-independent.  The action will often have additional 
symmetries, such as global SU(2) invariance; these depend entirely on the 
symmetries of the spin Hamiltonian.

We specialize to the Heisenberg Hamiltonian, $\hat{\cal H} = 
\sum_{<ij>}{\cal J}_{ij}\hat{\bf S}_{i} \cdot \hat{\bf S}_{j}$.
Under the Schwinger-boson transformation
$\hat{\bf S}_{i}\cdot \hat{\bf S}_{j}\mapsto$
:$\hat{B}_{ij}^{\dag}\hat{B}_{ij}$:$-\hat{A}_{ij}^{\dag}\hat{A}_{ij}$, 
where
$\hat{A}_{ij} = \frac{1}{2}(\hat{a}_{i}\hat{b}_{j} - \hat{b}_{i}\hat{a}_{j})$,
$\hat{B}_{ij}=\frac{1}{2}(\hat{a}_{i}^{\dag}\hat{a}_{j} + 
\hat{b}_{i}^{\dag}\hat{b}_{j})$,
and :$\hat{\cal O}$: is the normal-ordered form of $\hat{\cal O}$.
The bilinear operators $\hat{A}_{ij}$ and $\hat{B}_{ij}$ are invariant under
simultaneous SU(2) rotations of spins $i$ and $j$, so they may take on vacuum
expectation values even without spontaneous rotational symmetry breaking.
This makes them convenient order parameters on which to base a 
mean-field theory.  These operators are not, however, invariant under the U(1) gauge
transformations described earlier (since they couple physical and
unphysical states), so such theories retain some gauge degrees of 
freedom.  These degrees of freedom become important when
the order parameters are allowed to fluctuate around their average values.

We write the Hamiltonian in a form suitable for lattices with several 
spins per unit cell:
\begin{equation}
\hat{\cal H} = \sum_{{\bf x}n}{\cal J}_{n}
\hat{\bf S}_{{\bf x},\,\mu_{n}} \cdot
\hat{\bf S}_{{\bf x}+{\bf y}_{n},\,\nu_{n}}.
\end{equation}
Here ${\bf x}$ labels the unit cell, $\mu$ and 
$\nu = 1,2,\dots N_{\rm sites}$ 
are site labels within a unit cell, and $n = 1,2,\dots N_{\rm bonds}$ is a bond label.
We then formulate a mean-field theory with order parameters $\{A_{{\bf x}n}, B_{{\bf 
x}n}\}$ by introducing complex Hubbard-Stratonovich fields $\alpha$ 
and $\beta$, linearly coupling them to $A$ and $B$, and using them to 
integrate out the original boson fields $\phi$.  This yields a new
representation of the partition function,
\begin{equation}
\label{functionalZ2}
{\cal Z}(\beta) = 
\frac
{
\int{\cal D}\alpha{\cal D}\beta{\cal D}\lambda
\,\,
\delta[\partial_{\tau}\lambda]
\exp\left(
-{\cal A}_{\rm eff}[\alpha,\,\beta,\,\lambda]\right)
}
{
\int{\cal D}\alpha{\cal D}\beta{\cal D}\lambda
\,\,
\delta[\partial_{\tau}\lambda]
\exp\left(
-{\cal A}_{\rm HS}[\alpha,\,\beta]\right)
},
\end{equation}
where the effective and Hubbard-Stratonovich actions are
\begin{subequations}
\begin{eqnarray}
\label{effaction}
{\cal A}_{\rm eff}
&=& {\cal A}_{\rm HS} + {\rm Tr}\ln\hat{\rm M} +
\int_{0}^{\beta} \!\!\! d\tau \sum_{{\bf x}\mu} -2 i S_{\mu}\lambda_{{\bf x}\mu},
\\
{\cal A}_{\rm HS}
&=& \int_{0}^{\beta} \!\!\! d\tau  
\sum_{{\bf x}n}{\cal J}_{n}(\alpha_{{\bf x}n}^{*}\alpha_{{\bf x}n}
+
\beta_{{\bf x}n}^{*}\beta_{{\bf x}n}).
\end{eqnarray}
\end{subequations}
The denominator of (\ref{functionalZ2}) can be evaluated exactly, but 
it is convenient for normalization purposes to leave it as it stands.
The dynamical matrix $\hat{\rm M}[\alpha,\,\beta,\,\lambda]$, defined in the Appendix,
describes the propagation of Schwinger bosons coupled to the Hubbard-Stratonovich
fields.

As noted by Trumper {\it et al.}, the effective action 
(\ref{effaction})
has a gauge symmetry inherited from the boson action 
(\ref{action1})~\cite{trumper}.  The gauge-fixing functional,
currently given by $\delta[\partial_{\tau}\lambda]$, can
be chosen more or less arbitrarily without changing the value of
${\cal Z}$; this can be done with the Faddeev-Popov procedure.
However, while the exact value of ${\cal Z}$ is gauge-invariant, the
terms in a general perturbative expansion of it are not.
We will therefore retain the current gauge-fixing condition in what follows, but
observe that this quantitatively affects our results, because
the saddle-point expansion we use is not controlled by any 
gauge-invariant small parameter.  By contrast, mean-field theories which
take as order parameters either $\{A_{{\bf x}n}\}$ or $\{B_{{\bf x}n}\}$,
but not both, have natural SU(N) generalizations in which the 
saddle-point expansion is controlled by 1/N~\cite{auerbach}.

Next we look for stationary points of ${\cal A}_{\rm eff}$ with
respect to its arguments.  We consider only static and 
translationally invariant solutions of the form
$\alpha_{{\bf x}n} = \alpha_{{\bf x}n}^{*} = \alpha_{n}$,
$\beta_{{\bf x}n} = -\beta_{{\bf x}n}^{*} = \beta_{n}$,
and
$\lambda_{{\bf x}\mu} = -i\Lambda_{\mu}$, where $\alpha$, $\beta$, and $\Lambda$ are real, and
we work in the zero-temperature limit.  With these restrictions we
can derive the desired rotationally invariant mean-field equations.
Collectively they take the variational form $\delta E = 0$, where the energy per unit cell
is given by
\begin{eqnarray}
\label{MFenergy}
\frac{1}{N}E(\alpha, \beta, \Lambda) &=& 
\sum_{n}{\cal J}_{n}(\alpha_{n}^{2} - \beta_{n}^{2})
-
\sum_{\mu}(2 S_{\mu} + 1)\Lambda_{\mu} \nonumber\\
&&+
\frac{1}{N}\sum_{{\bf k}\mu} \omega_{{\bf k}\mu},
\end{eqnarray}
and the quasiparticle energies $\{\omega_{{\bf k}\mu}\}$ are defined
in the Appendix.  This represents a set of $2 N_{\rm bonds} + N_{\rm 
sites}$ coupled nonlinear equations, which must be solved numerically
for the mean-field parameters $\alpha^{(0)}$, $\beta^{(0)}$, and $\Lambda^{(0)}$.
(There are generally several gauge-equivalent saddle points, differing 
only in the signs of $\alpha$ and $\beta$; which one of these is chosen
as the mean-field solution is irrelevant.)
The resulting boson propagator $\hat{\cal G}^{(0)} = (\hat{\rm M}^{(0)})^{-1}$
enters into the calculation of fluctuation corrections.
The ground-state energy per unit cell can be rewritten as $\frac{1}{N}E^{(0)} = 
\sum_{n}{\cal J}_{n}\left((\beta_{n}^{(0)})^{2} - (\alpha_{n}^{(0)})^{2}\right)$.

Finally, we include fluctuations around the saddle point in
the usual way, by expanding the action in a power series:
\begin{equation}
{\cal A}_{\rm eff} = {\cal A}^{(0)} + 
\frac{1}{2} \Psi^{\dag}
\left.
\frac{\partial^{2}{\cal A}_{\rm eff}}{\partial\Psi^{\dag}\partial\Psi}
\right\arrowvert_{(0)}
\Psi + {\cal O}(\Psi^{3}),
\end{equation}
where the fluctuation fields $\Psi^{\dag} =
(\alpha^{*}, \, \alpha, \, \beta^{*}, \, \beta) -
(\alpha^{*}, \, \alpha, \, \beta^{*}, \, \beta)^{(0)}$ do not include $\lambda$ due to our choice of
gauge.  The matrix of second derivatives, which is block-diagonal in
reciprocal space, determines the Gaussian correction to the ground-state energy:
\begin{equation}
\label{e1no1}
E^{(1)} = 
\int\displaylimits_{-\infty}^{+\infty}\frac{d\omega}{4\pi}\sum_{\bf 
k}{\rm Tr}\ln\left(\hat{\openone} - \hat{\rm D}({\bf k},\,\omega)\right),
\end{equation}
where
\begin{equation}
{\rm D}_{a m, b n}({\bf k},\omega) =
{\cal J}_{m}^{-1} {\rm Tr}\left(
\hat{\cal G}^{(0)}
\frac{\partial\hat{\rm M}}{\partial\Psi^{\dag}_{{\bf k}\omega, 
a m}}
\hat{\cal G}^{(0)}
\frac{\partial\hat{\rm M}}{\partial\Psi_{{\bf k}\omega, b n}}
\right).
\end{equation}
These expressions, together with the details provided in the Appendix, 
represent our main analytical results.  We now discuss the numerical
results for a particular lattice.

\section{Shastry-Sutherland lattice}

\begin{figure}
\begin{center}
\includegraphics[width=3.25 in]{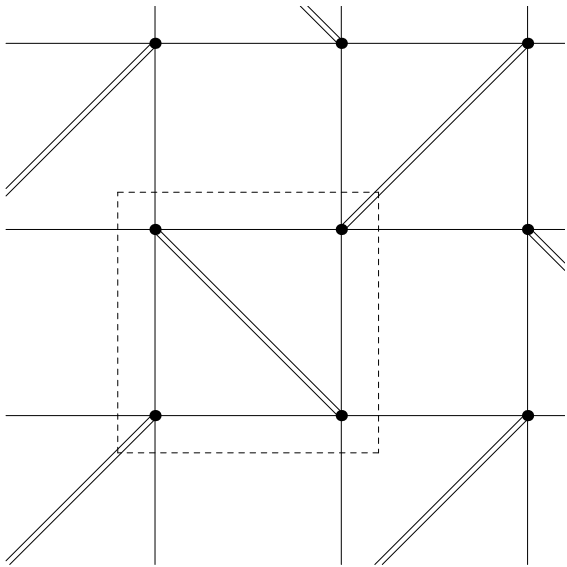}
\end{center}
\caption{\label{fig1}Shastry-Sutherland lattice.  Single- and double-line bonds
have strengths $J_{1}$ and $J_{2}$ respectively.  The dashed square
contains one unit cell.}
\end{figure}

The Shastry-Sutherland model is a frustrated 2D antiferromagnet with 
bonds of two different strengths, as shown in Fig.~\ref{fig1}.  The relevant 
physical parameters are the ratio $g\equiv{J_{2}/J_{1}}$ and the spin 
magnitude $S$.  In the classical limit ($S\rightarrow\infty$)
the ground state has N\'eel order for $g \le 2$ 
and helical long-range order otherwise.  For finite $S$ and 
sufficiently large $g$, the exact quantum ground state is known: it is fully
dimerized, with the two spins on each ${J_{2}}$-bond in the
singlet state~\cite{shassuth}.  (This is often referred to as a spin-liquid phase
due to the lack of long-range order.)  The energy of this
state is always $-2S(S+1)J_{2}$ per unit cell.
Moreover one expects the N\'eel
order present at $g=0$ to persist up to some positive $g_{c}(S)$.
We would like to clarify to what extent quantum fluctuations
melt the classical spiral ordering.
Our main interest is in the case $S=1/2$.  This is the physically
relevant case, and the existence of a spiral phase is in the greatest
doubt here.  A previous Schwinger-boson calculation at the mean-field level
showed a spiral phase in the range $1.1 \le g \le 1.65$~\cite{albmila}.
This result has not been confirmed by exact diagonalization, since the spiral 
states are only energetically favorable on large lattices.  
High-order series expansions, starting from both $g=\infty$ and 
$g=0$, have been used to locate accurately the energetic crossover between the N\'eel and 
spin-liquid phases~\cite{series}; however, these calculations provide no evidence 
for or against an intermediate spiral phase.

When describing helical phases it is common to use ``twisted'' 
periodic boundary conditions, where spins along one cluster boundary
are rotated with respect to those on the opposite 
boundary~\cite{twist1,twist2}.
This procedure breaks rotational invariance and complicates
the analysis somewhat, so we do not use it here.  The price we pay is 
that helices with small pitch cannot be accommodated by small clusters 
with untwisted boundary conditions.  We found that helical mean-field solutions
exist for clusters with as few as 24 sites, suggesting that finite-size effects 
probably do not seriously alter our results on larger clusters.

We used the Schwinger-boson method to calculate the ground state energy
for $S=1/2$.
The mean-field equations can be solved numerically for the full range 
of $g$.  For small $g$, the solution with lowest energy
is purely antiferromagnetic ($\alpha\neq 0$, $\beta=0$) on the strong ${J_{1}}$-bonds and purely 
ferromagnetic ($\beta\neq 0$, $\alpha = 0$) on the weaker ${J_{2}}$-bonds.
This solution is unique and has ordering 
wavevector ${\bf Q}=(\pi,\pi)$; we identify it as the N\'eel 
state.  When $g$ is increased, new solutions with ordering wavevectors ${\bf 
Q}=(\pi\pm\delta q,\pi)$ and ${\bf Q}=(\pi,\pi\pm\delta q)$ appear 
through a bifurcation of the N\'eel state; we 
identify these solutions as helical states with pitch along
the $x$-axis and $y$-axis respectively.  The helical states are 
degenerate for square lattices.  Finally, for sufficiently large $g$, the
energy of the helical states exceeds that of the spin-liquid state, and
the ground state becomes disordered.

\begin{figure}
\begin{center}
\includegraphics[width = 3.25 in]{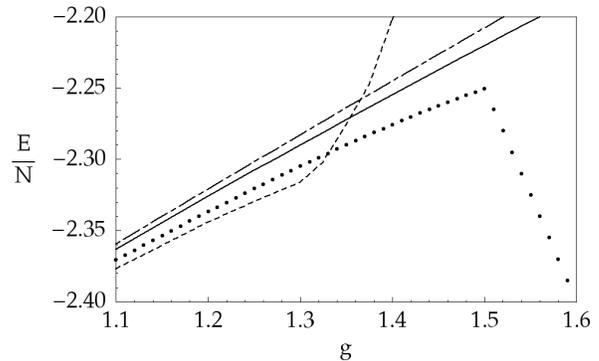}
\end{center}
\caption{\label{fig2}Ground-state energy, small cluster.  The energy per unit cell
is plotted, in units of $J_{1}$, as calculated by (from top to 
bottom at figure left) perturbation theory, Schwinger-boson theory (mean field), exact diagonalization,
and Schwinger-boson theory (Gaussian order).}
\end{figure}

The results for a small cluster (16-site, $4\times 4$) are shown in 
Fig.~\ref{fig2}.  The N\'eel state does not undergo any bifurcation here, due to the 
small size of the cluster.  The Schwinger-boson mean-field estimate exceeds 
the exact ground state energy by a small and roughly constant amount over the 
full range of $g$.  A second-order perturbative calculation around 
$g=0$ (that is, around the N\'eel state) yields 
similar results in the displayed range (although naturally perturbation theory
is more accurate near $g=0$).  Finally, the inclusion of Gaussian fluctuations 
in the Schwinger-boson calculation improves the accuracy, but for $g\lesssim 
1.3$ only: for larger $g$ the results deteriorate rapidly.  

\begin{figure}
\begin{center}
\includegraphics[width = 3.25 in]{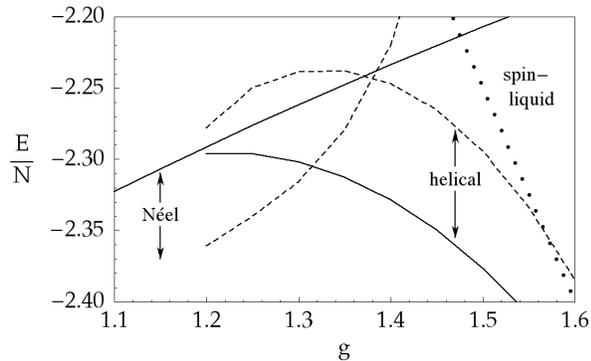}
\end{center}
\caption{\label{fig3}Ground-state energy, large cluster.  As in the previous 
figure, the solid curves 
represent Schwinger-boson mean-field results, while the dashed curves are corrected for
Gaussian fluctuations.  These energies are given for both N\'eel and
helical phases, as indicated.  The exact energy of the spin-liquid 
phase is also shown (dotted line).}
\end{figure}

In Fig.~\ref{fig3} we show the mean-field and Gaussian-order results 
for a larger, rectangular cluster (64-site, $16\times 4$), for which
exact results are unavailable (except for the spin-liquid state).  In this
case the N\'eel state bifurcates at $g \simeq 1.2$ into a helical state
with pitch along the $x$-axis (the lower dashed curve).  The energy of
the N\'eel state is corrected by 
fluctuations just as in the smaller cluster: it is shifted
downward by a small amount until $g \simeq 1.3$, after which the
computed correction becomes large and positive.  Extrapolating from 
the small-cluster results, it seems probable that the upswing is again
spurious, and that the true energy is bracketed by the mean-field and
Gaussian-order results.  The Gaussian correction to the energy of the 
helical state is uniformly positive.

We also checked the mean-field and Gaussian-order results for still
larger clusters, at selected values of $g$.  The bifurcation point
occurs at slightly smaller $g$ on larger clusters, approaching $g=1.1$
in the thermodynamic limit, as seen in Ref.~\cite{albmila}.  
The nature of the corrections does not appear to change.  Our
conclusion is that the Shastry-Sutherland model with $S=1/2$ has a
helical ground state, over a range of $g$ narrower than that suggested by
mean-field Schwinger-boson theory.  The range of stable wavevectors 
for spiral ordering is correspondingly reduced: we estimate that the
helical state is stable for $\delta q \lesssim \pi/3$, whereas the
mean-field theory predicts stability for $\delta q \lesssim 3\pi/8$.
These conclusions are necessarily tentative.

\section{Remarks}

From a practical point of view, these results are not very satisfactory.
Although the inclusion of Gaussian fluctuations improves the agreement
between Schwinger-boson theory and exact diagonalization in some
regions of parameter space, the improvement is not dramatic.  
Moreover, the correction is not uniformly good; it gives little
reliable information about the region of interest, where the N\'eel and helical states
have comparable energies.

The results can be interpreted in two ways.
It may be that the very good agreement between the mean-field and
exact ground-state energies is simply fortuitous; the
partition function may not be dominated by contributions from the
neighborhood of the mean-field solution.  In that case no
refinement of the present saddle-point expansion will yield
better results.  This is a real possibility; however, since the
Schwinger-boson mean-field approach also gives accurate results
on a number of other systems, it seems unlikely.

It also may be that we have failed to treat the gauge degrees of
freedom properly.  As noted earlier, the choice of gauge-fixing
condition has a quantitative effect on the order-by-order results
of the saddle-point expansion.\footnote{\label{foot1}The authors of Ref.~\cite{trumper}
indicate that their results for a different lattice are independent
of whether our gauge or the background gauge is chosen.  We were
unable to reproduce this coincidence for the Shastry-Sutherland
model.}
We worked in a simple but otherwise 
arbitrary gauge.  It would be preferable
to control the expansion in a gauge-invariant way.  This
should be possible along the lines of Ref.~\cite{auerbach} 
(and references therein): allow the number of boson ``flavors'' to
become large, say N, and generalize the fields $A$,
$B$,
and ${\delta n}$ in a natural 
way to include these new flavors.  When this is done correctly, the
effective action for $\alpha$, $\beta$, and $\lambda$ will be 
proportional to N; the saddle-point expansion will then yield an
asymptotic series in 1/N, each term of which is gauge-invariant.
This program would put the present work on a sounder theoretical 
footing, by determining in what limit its results become exact.
The key step is to generalize $A$ and $B$ simultaneously; this
remains a challenge for the future.

\begin{acknowledgments}
This work was made possible in part by the kind support of Cornell
University.  The author was also supported in part by an ITP Graduate
Fellowship from the Institute for Theoretical Physics at the University
of California, Santa Barbara.  That funding was through the NSF under
Grant No. PHY89-04035.  The author would like to thank Guang-Hong
Chen for his collaboration during the early stages of this project.
\end{acknowledgments}

\appendix

\section{Details of calculation}
\label{sec:appendix}

This appendix contains details about some aspects of the calculation 
which, while extraneous to the main body of the paper, are included
for the sake of completeness.

The boson action (which was integrated out) has the form $(\phi_{a}^{*} 
\,\phi_{b})\cdot\hat{\rm M}\cdot(\phi_{a}\,\phi_{b}^{*})^{\rm T}$; i.e., it is 
anomalous.  The dynamical matrix $\hat{\rm M}[\alpha,\,\beta,\,\lambda]$ has $2\times2$ submatrices
\begin{eqnarray}
\label{Mmatrix}
&&\!\!\!\!(\hat{\rm M})_{{\bf k}\omega\mu,{\bf k}'\omega'\mu'} =
\beta N \delta_{{\bf k}\omega\mu,{\bf k}'\omega'\mu'}
\begin{pmatrix}
	i\omega e^{i\epsilon\omega} & 0 \cr 0 & -i\omega e^{-i\epsilon\omega}
\end{pmatrix}
\nonumber\\
&&+
\left(
i\delta_{\mu\mu'}\tilde\lambda_{\mu} +
\sum_{n}
(\hat{\rm F}^{+}_{n})_{{\bf k}\mu,{\bf k}'\mu'}
\left(
-i\widetilde{{\rm Im}\beta_{n}}\right)
\right)\hat{\openone}\nonumber\\
&&+
\sum_{n}
(\hat{\rm F}^{-}_{n})_{{\bf k}\mu,{\bf k}'\mu'}
\left(
- \widetilde{{\rm Re}\beta_{n} }\hat\sigma_{z}
-i\widetilde{{\rm Re}\alpha_{n}}\hat\sigma_{y}
+i\widetilde{{\rm Im}\alpha_{n}}\hat\sigma_{x}
\right),\nonumber\\
\end{eqnarray}
where all fields are Fourier-transformed, with arguments
$({\bf k}-{\bf k}',\,\omega - \omega')$ suppressed, and
\begin{equation}
(\hat{\rm F}^{\pm}_{n})_{{\bf k}\mu,{\bf k}'\mu'} =
\frac{1}{2}{\cal J}_{n}
\left(
\delta_{\mu\mu_{n}}\delta_{\mu'\nu_{n}}e^{i{\bf k}'\cdot{\bf y}_{n}} \pm
\delta_{\mu'\mu_{n}}\delta_{\mu\nu_{n}}e^{-i{\bf k}\cdot{\bf y}_{n}}
\right).
\end{equation}
The differing convergence factors $e^{\pm i\epsilon\omega}$ in 
the first term of $\hat{\rm M}$ arise from the anomalous nature of
the action.  They must be taken into account properly to obtain the correct
mean-field energy, Eq.~(\ref{MFenergy}), and to evaluate 
the Gaussian correction given by Eq.~(\ref{e1no1}).  However,
in the latter equation only the first $(n=1)$ term from
the expansion ${\rm Tr}\ln({\hat\openone}-{\hat{\rm D}}) = -\sum_{n}\frac{1}{n}{\rm Tr}\,{\hat{\rm D}}^{n}$
needs these factors to force convergence of the $\omega$-integral, and the contribution to $E^{(1)}$ 
from this term can be shown to equal $\frac{1}{2}E^{(0)}$.
For numerical work, then, it is simplest to rewrite the Gaussian correction as
\begin{equation}
E^{(1)} = \frac{1}{2}E^{(0)} + 
\int\displaylimits_{-\infty}^{+\infty}\frac{d\omega}{4\pi}\sum_{\bf k}
\left({\rm Tr}\,\hat{\rm D} + {\rm Tr}\ln(\hat{\openone} - \hat{\rm D})\right)
\end{equation}
and disregard the convergence factors.

The mean-field dynamical matrix is block-diagonal:
\begin{equation}
\frac{1}{\beta N}\hat{\rm M}^{(0)}({{\bf k},\omega}) =
i\omega{\hat\openone}\otimes{\hat\sigma}_{z}
+ {\hat A}({\bf k})\otimes{\hat\openone} + {\hat B}({\bf k})\otimes{\hat\sigma}_{y},
\end{equation}
where ${\hat A}$ and ${\hat B}$ are Hermitian matrices
obtained from the general expression for $\hat{\rm M}$.
This can be diagonalized using
a nonunitary (generalized Bogoliubov) transformation: there is a
matrix $\hat{\cal S}({\bf k}) = {\hat G}({\bf k})\otimes{\hat\openone} +
{\hat H}({\bf k})\otimes{\hat\sigma}_{y}$ such that
$\hat{\cal S}^{\dag} = ({\hat\openone}\otimes{\hat\sigma}_{z})\hat{\cal 
S}^{-1}({\hat\openone}\otimes{\hat\sigma}_{z})$ and
\begin{equation}
\frac{1}{\beta N}(\hat{\cal S}^{\dag}\hat{\rm M}^{(0)}\hat{\cal S})_{\mu\mu'} =
\delta_{\mu\mu'}
\begin{pmatrix}
	i\omega + \omega_{{\bf k}\mu} & 0 \cr 0 & -i\omega + \omega_{{\bf k}\mu}
\end{pmatrix}
.
\end{equation}
The quasiparticle energies $\{\omega_{{\bf k}\mu}\}$ are the
square roots of the eigenvalues of $({\hat A}+{\hat B})({\hat 
A}-{\hat B})$; these eigenvalues will be real and positive
when $\Lambda$ is large enough, and the mean-field solutions must
be sought in that region.  The diagonalizing matrix and the 
quasiparticle energies are analytically tractable for one-
and two-site unit cells; in any case they are easy to obtain
numerically.  The boson propagator is then
\begin{eqnarray}
&&\!\!\!\!\beta N\hat{\cal G}^{(0)}_{\mu\mu'}({\bf k},\omega) = \nonumber\\
&&\;\;\;\hat{\cal S}_{\mu\nu}({\bf k})
\begin{pmatrix}
	(i\omega + \omega_{{\bf k}\nu})^{-1} & 0 \cr 0 & 
	(-i\omega + \omega_{{\bf k}\nu})^{-1}
\end{pmatrix}
\hat{\cal S}^{\dag}_{\nu\mu'}({\bf k}). \nonumber\\
\end{eqnarray}
Finally, the mean-field derivatives of $\hat{\rm M}$ with
respect to the fluctuation fields are field- and
frequency-independent, and can be read directly from
Eq.~(\ref{Mmatrix}).


\end{document}